\documentclass[11pt]{article}

\usepackage[margin=1in]{geometry}
\usepackage{times}
\usepackage{microtype}
\usepackage{setspace}
\usepackage{titlesec}
\titleformat{\section}{\large\bfseries}{\thesection}{1em}{}
\titleformat{\subsection}{\normalsize\bfseries}{\thesubsection}{1em}{}

\usepackage[T1]{fontenc}
\usepackage{graphicx,adjustbox,subfigure,xcolor,tikz}
\usetikzlibrary{positioning}
\usepackage{amsmath,amssymb}
\usepackage{multirow,array,enumitem}
\usepackage{tabularx}
\usepackage{pifont}
\usepackage{hyperref}
\usepackage{natbib}
\usepackage{booktabs}
\usepackage{balance}

\title{Pre-training vs. Fine-tuning: A Reproducibility Study on\\Dense Retrieval Knowledge Acquisition}

\author{
	Zheng Yao$^{1}$ \\
	\small The University of Queensland, Australia \\
	\small \texttt{zheng.yao1@student.uq.edu.au} \\
	\and
	Shuai Wang$^{1}$ \\
	\small The University of Queensland, Australia \\
	\small \texttt{shuai.wang2@uq.edu.au} \\
	\and
	Guido Zuccon$^{1}$ \\
	\small The University of Queensland, Australia \\
	\small \texttt{g.zuccon@uq.edu.au} 
}

\date{}

\begin{document}
	\maketitle
	
	\begin{abstract}
		Dense retrievers utilize pre-trained backbone language models (e.g., BERT, LLaMA) that are fine-tuned via contrastive learning to perform the task of encoding text into sense representations that can be then compared via a shallow similarity operation, e.g. inner product. Recent research has questioned the role of fine-tuning vs. that of pre-training within dense retrievers, specifically arguing that retrieval knowledge is primarily gained during pre-training, meaning knowledge not acquired during pre-training cannot be sub-sequentially acquired via fine-tuning. We revisit this idea here as the claim was only studied in the context of a BERT-based encoder using DPR as representative dense retriever. We extend the previous analysis by testing other representation approaches (comparing the use of CLS tokens with that of mean pooling), backbone architectures (encoder-only BERT vs. decoder-only LLaMA), and additional datasets (MSMARCO in addition to Natural Questions). Our study confirms that in DPR tuning, pre-trained knowledge underpins retrieval performance, with fine-tuning primarily adjusting neuron activation rather than reorganizing knowledge. However, this pattern does not hold universally, such as in mean-pooled (Contriever) and decoder-based (LLaMA) models. We ensure full reproducibility and make our implementation publicly available at \url{https://github.com/ielab/DenseRetriever-Knowledge-Acquisition}.
	\end{abstract}
	
	\textbf{Keywords:} Dense Retrieval, Fine-tuning, Pre-training, Knowledge Acquisition
	
	\section{Introduction}
\label{sec:intro}

Dense retrievers are embedding models based on pre-trained language models that are fine-tuned, typically via contrastive learning, to improve retrieval effectiveness~\cite{karpukhin2020dense,trabelsi2021neural,lin2021pretrained,tonellotto2022lecture,fan2022pre,zhao2024dense,wang2022neural}. Dense retrieval research has primarily considered improvements in training matters (e.g., loss function~\cite{zhang2021lossfunction} and negative sampling~\cite{chen2020negativesampling,wang2023balanced,hofstatter2021efficiently}), representational power (e.g., CLS token~\cite{liu2019clstoken}, mean pooling~\cite{sun2020meanpooling}), and underlying language modelling backbones (e.g., encoder-based~\cite{devlin2019bert, wang20242d} and decoder-based~\cite{ma2024fine,li2024llama2vec,zhuang2024promptreps}). However, little attention has been put into fundamentally understanding the impact that dense retrieval fine-tuning has on the information flow and distribution within the language model backbone (i.e. the trained Transformer). For example the answers to questions such as: \textit{``Do dense retrievers learn new knowledge during dense retrieval fine-tuning?''} and, \textit{``Does this fine-tuning modify a pre-trained model's internal knowledge structure?''} are yet not well understood. 
%
Answering these questions is important as such answers could have important implications in the training and usage of dense retrievers. For example, if dense retrieval fine-tuning does learn new knowledge that was not learn by pre-training, then we could rather directly fine-tune on the retrieval tasks a non pre-trained model (i.e. a vanilla Transformer), potentially achieving high effectiveness without the costly pre-training phase. 
This would be in line with suggestions from researchers to create language modelling training methods (and consequently backbones) specifically designed for information retrieval, in place of the current approach of adapting language models trained on word prediction tasks~\cite{chang2020pre-training,ma2021pre,ma2021prop,lin2021pretrained,zhuang2024starbucks}. Similarly, if fine-tuning cannot modify or update the internal knowledge of a pre-trained model, a more deliberate and carefully designed pre-training process may be necessary to enhance the trustworthiness of dense retrieval models.

The recent work of \citeauthor{reichman2024dense} has attempted to shed light on this matter~\cite{reichman2024dense}. Through experimenting with a pre-trained BERT backbone and a version of that backbone that has further undergone DPR fine-tuning~\cite{karpukhin2020dense}, they suggest that fine-tuning may not introduce new knowledge but instead it \emph{decentralizes} existing representations to improve retrieval effectiveness. If this claim holds, the upper-bound effectiveness of dense retrieval systems may be inherently constrained by the choice of pre-trained backbone models. This claim is based on analyses they carried out that used linear probing, neuron activation studies, and knowledge editing. 

The original paper argues that DPR fine-tuning does not introduce new knowledge but rather reorganizes or decentralizes existing knowledge by increasing activations in intermediate layers. This implies -- the original authors argue -- that the pretrained model already contains the knowledge, and effective retrieval of new knowledge would require pretraining, not just fine-tuning\footnote{The original paper also considers previous methods for knowledge editing and their impact on dense retriever effectiveness. Our investigation is focused on understanding knowledge acquisition and decentralization, and thus we do not consider the original knowledge editing experiments.}.

In this paper, we systematically investigate \citeauthor{reichman2024dense}'s hypothesis by reproducing and extending their analyses of knowledge representation in dense retrievers. We expand the experimental scope from a single dataset (Natural Questions) to include MS MARCO, ensuring broader applicability of findings. Additionally, while the original investigation only focused on DPR, which utilized the CLS token as embedding representation, we also consider a different embedding representation method, mean-pooling, which is used by the state-of-the-art Contriever approach~\cite{izacard2022unsupervised}. Finally, we also consider a decoder-only generative LLM backbone in place of the BERT-base-uncased model used by \citeauthor{reichman2024dense}, investigating the behaviour of Repllama~\cite{ma2024fine}, which relies on the Llama LLM; this also  provides us with the opportunity to investigate one more representation method along with CLS token and mean-pooling: the use of the EOS token.
In our work we focus on two key analysis tasks: (1) \textbf{linear probing}, which examines layer-wise discriminative capacity to determine how fine-tuning reshapes model representations, and (2) \textbf{integrated gradient analysis}, which tracks neuron activation patterns to assess knowledge decentralization effects.



Through comprehensive experiments, we  observe the same trend identified by \citeauthor{reichman2024dense} when comparing the BERT backbone and its fine-tuned DPR model on both the NQ and MS MARCO datasets. However, this trend changes when considering the Contriever fine-tuning recipe (and mean-pooling instead of CLS token), and so it does also when changing to a decoder-only backbone model and EOS token representations. Our results suggest that the decentralization effect observed by \citeauthor{reichman2024dense} for DPR may not generalize uniformly across different architectures and embedding strategies, highlighting the need for further investigation into model-specific adaptation dynamics.

	\section{Reproduction Challenges}\label{sec_repro_challenges}

We encountered several challenges in reproducing the work of \citeauthor{reichman2024dense}; these were primarily due to lack of a reference codebase that implements their methods, and lack of details, mostly related to data processing. Next, we further detail these challenges and how we tackled them; in the subsequent sections, we then describe in-depth the analysis methodology and the experimental results.

\textbf{Datasets Challenges.} The original paper relied on the Natural Questions (NQ) dataset for experimentation. A key aspect of the linear probing methodology used in the experiments is the availability of hard negative labels for queries/passages in NQ -- in particular \citeauthor{reichman2024dense} built probes with up to four hard negative passages. However, they did not report how these hard negative labels were obtained. The original NQ dataset does not contain negative labels. Multiple researchers have extended NQ with additional labels, including (hard) negative labels. For example, the data set made available at \url{https://huggingface.co/datasets/tomaarsen/natural-questions-hard-negatives} provides five hard negative passages for each NQ query; labels have been mined using the \texttt{all-MiniLM-L6-v2} model\footnote{\url{https://huggingface.co/sentence-transformers/all-MiniLM-L6-v2}.}. However, it is highly unlikely that this specific dataset was used by \citeauthor{reichman2024dense} as its release date postdates that of the their paper. We opted for using the hard negative labels provided as part of extensions to the original DPR paper~\cite{karpukhin2020dense} and made available at \url{https://github.com/facebookresearch/DPR}. A drawback with using this dataset with hard negatives is that it does not provide exactly four hard negatives for all NQ queries: some queries can have more than four, while some others can have none. In Section~\ref{sec:experiment-setup} we further describe how we adapted this data set for our experiments.

	
\textbf{Linear Probing Training Challenges.} Another key aspect in the analysis based on linear probing is the need to train the linear probing layer. \citeauthor{reichman2024dense} did not detail how this training was performed, nor did they mention the structure of the probing layer or report the training hyper-parameters. In Section~\ref{sec:methodology-task2} we detail the structure of the probing layer we employed and the corresponding training procedure; our hyper-parameters and other training details are summarised in  Table~\ref{table:training_parameters}.
 
\begin{table}[t]\centering
	\small
	\caption{Hyper-parameters and infrastructure used for the training of the linear probing layer.}
	\label{table:training_parameters}
	\begin{tabular}{lc} 
		\toprule
		\bfseries Hyper-parameter & \bfseries Assignment  \\
		\midrule
		train-validation ratio& 0.99 \\
		random seed & 42 \\
		learning Rate & 1e-4\\
		batch size & 32768 \\
		maximum epochs& 30 \\
		GPU specs& 1 x H100 96GB \\
		\bottomrule
	\end{tabular}
\end{table}

\textbf{Knowledge Decentralization Investigation Challenges.} The original paper investigated how knowledge neurons in the models are differently activated when considering the pre-trained vs. the fine-tuned model. 
Identifying knowledge neurons requires selecting a target task, as neuron attribution is assessed by observing changes in task performance when adjusting the neuron weights of each layer. In dense retrieval, two potential tasks could be considered: (1) evaluating retrieval effectiveness using traditional IR metrics, e.g., nDCG@10, or (2) measuring the impact of weight adjustments on text embeddings by analyzing vector changes. \citeauthor{reichman2024dense}  did not clearly specify which approach was used to determine neuron weights. In our experiments, we adopted the second approach. We arrived at this choice by examining the relevant plots in the original paper (Figures 1 and 2 in \cite{reichman2024dense}). In these figures, the magnitude scale of the results largely differs depending whether the DPR-question encoder is used (scale ranging from 0 to below 6,000) or DPR-context is used (scale taking values above 60,000), implying that different sets of query-document pairs were used depending on the encoder. 

	\section{Research Questions}

To guide our reproducibility efforts and the investigation of the empirical results, we formulate four research questions:

\begin{itemize}
	\setlength\itemsep{3pt}
\item[\bf{RQ1}] \textbf{(same settings):} Can we reproduce the findings of \citeauthor{reichman2024dense} using the same dataset and model configurations?

\item[\bf{RQ2}] \textbf{(different dataset):} Do findings generalize to other datasets?

\item[\bf{RQ3}] \textbf{(different recipe and representation):} Do findings obtained when using DPR fine-tuning and CLS token based representations apply to the Contriver fine-tuning recipe and mean pooling based representations?

\item[\bf{RQ4}] \textbf{(different backbone and representation):} Do findings obtained for the BERT backbone transfer to decoder-based backbones?
\end{itemize}
	\section{Embedding Knowledge Consistency}

We first investigate the consistency of embedding knowledge between dense retrievers and their untrained backbone counterparts. Specifically, we aim to determine whether fine-tuned dense retrieval models encode knowledge in a manner consistent with their pre-training initialization. To assess this, we conduct a linear probing experiment following the original methodology proposed by~\citet{reichman2024dense}. Linear probing involves training a simple linear classifier on embeddings extracted from different layers of a model to evaluate whether these embeddings encode discriminative information for performing a task for which they have not been trained. The task considered by \citeauthor{reichman2024dense} is that of distinguishing relevant and irrelevant documents with respect to a query, i.e. a classification task (this is different from the dense retrieval task, where the training is instead related to establishing the similarity between a query and a passage). If a linear classifier achieves high accuracy, it suggests that the embeddings contain structured and meaningful knowledge. We describe the details of this approach below.

\begin{figure*}[ht!]
  \centering
  \includegraphics[width=\textwidth]{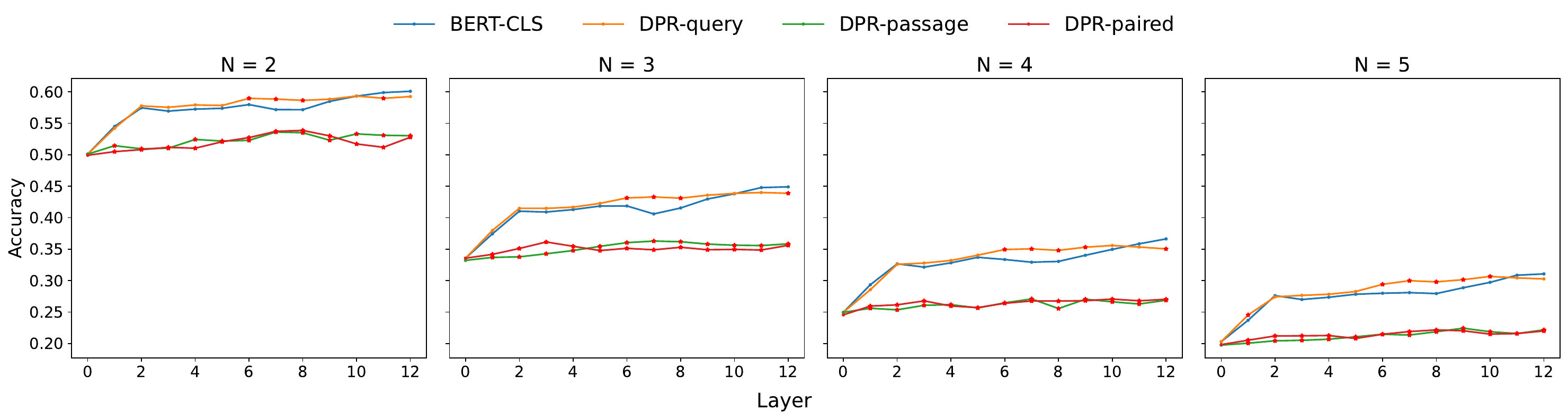}
  \caption{Results for linear probing experiments on the NQ dataset. The results are based on four different configurations for the number of passages in the probe ($N=2,3,4,5$), each represented by a sub-plot. The x-axis indicates the number of layers used for embedding learning. A two-tailed t-test ($p<0.05$) with Bonferroni correction between each DPR model and the BERT-base-uncased backbone model is marked with *; \citeauthor{reichman2024dense} did not perform statistical significance analysis so this aspect cannot be cross-checked.}
  \label{fig:cls_image}
\end{figure*}

\begin{figure*}[ht!]
  \centering
  \includegraphics[width=\textwidth]{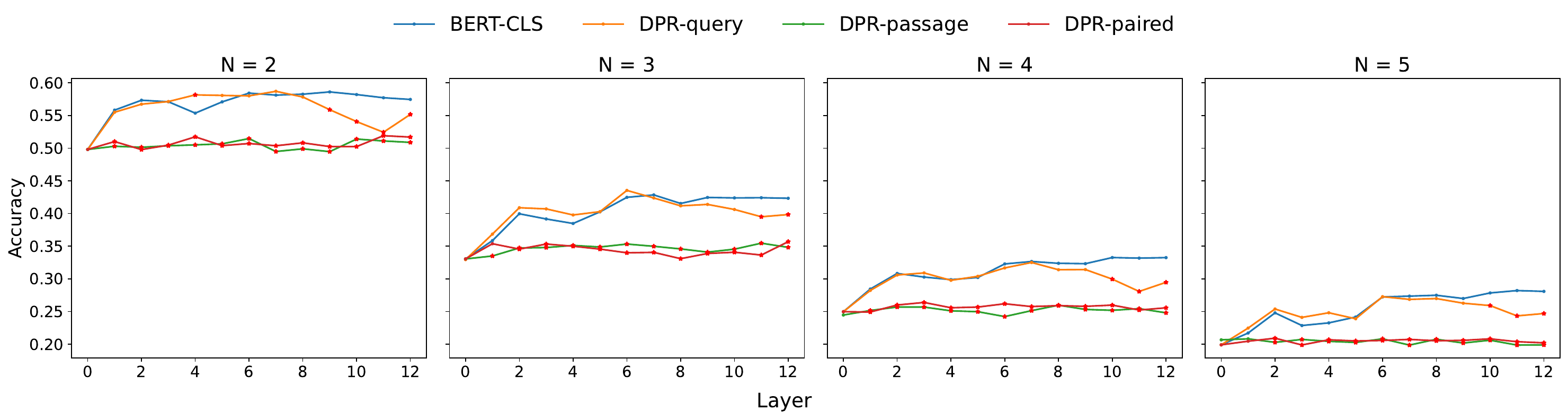}
  \caption{Generalization of linear probing experiments to MS MARCO. The results are based on four different configurations for the number of passages in the probe ($N=2,3,4,5$), each represented by a sub-plot. The x-axis indicates the number of layers used for embedding learning. A two-tailed t-test ($p<0.05$) with Bonferroni correction between each DPR model with bert-based-unacsed backbone model is marked with *. }
  \label{fig:cls_msmarco}
\end{figure*}

\subsection{Methodology: Linear-Probing Training}
\label{sec:methodology-task2}
\label{sec:methodology}

The linear probing experiment consists of two main stages: (1) pre-computation of embeddings and (2) training and evaluation of a linear classifier.

\paragraph{Stage 1: Pre-computation of Embeddings.} 
We extract embeddings from both fine-tuned dense retrieval models and their untrained backbone models. For a given input sequence $x$, let $H^{(\ell)}(x) \in \mathbb{R}^{T \times d}$ denote the hidden representation at layer $\ell$, where $T$ is the sequence length and $d$ is the embedding dimension. The final representation $\mathbf{e}^{(\ell)}(x)$ is computed as $\mathbf{e}^{(\ell)}(x) = f(H^{(\ell)}(x))$, 
where $f(\cdot)$ is a model-specific transformation. We consider three models:
\begin{itemize}[leftmargin=6pt]
    \item \textbf{DPR \cite{karpukhin2020dense}:} This is the only dense retrieval model considered by the original work of \citeauthor{reichman2024dense}. DPR uses the CLS token representation:
    \begin{equation}
    \mathbf{e}^{(\ell)}(x) = H^{(\ell)}_{0}(x).
    \end{equation}
    \item \textbf{Contriever \cite{izacard2022unsupervised}:} Uses mean pooling over all token representations:
    \begin{equation}
    \mathbf{e}^{(\ell)}(x) = \frac{1}{T} \sum_{t=1}^{T} H^{(\ell)}_{t}(x).
    \end{equation}
    \item \textbf{RepLlama \cite{ma2024fine}:} Uses the last token (EOS) representation:
    \begin{equation}
    \mathbf{e}^{(\ell)}(x) = H^{(\ell)}_{T-1}(x).
    \end{equation}
\end{itemize}

\paragraph{Stage 2: Training and Evaluating a Linear Classifier.} 
After extracting embeddings, we train a linear classifier to distinguish between relevant and irrelevant documents. The classifier takes the query embedding $\mathbf{q} \in \mathbb{R}^d$ and multiple passage embeddings $\mathbf{p}_1, \dots, \mathbf{p}_N \in \mathbb{R}^d$, where exactly one passage is labeled as relevant, while the others are labelled as hard negative. We construct a concatenated feature vector $\mathbf{z} = \left[\mathbf{q}; \mathbf{p}_1; \dots; \mathbf{p}_N\right] \in \mathbb{R}^{(N+1)d}$, 
which is projected through a learnable linear transformation $\mathbf{u} = \mathbf{W}\mathbf{z} + \mathbf{b}$, 
where $\mathbf{W} \in \mathbb{R}^{N \times (N+1)d}$ and $\mathbf{b} \in \mathbb{R}^{N}$.

The classifier is trained using a cross-entropy loss. Evaluation is based on classification accuracy, measuring the fraction of cases where the passage identified as relevant by the classifier is the ground-truth relevant passage. This methodology allows us to systematically compare the representational quality of fine-tuned dense retriever models with their untrained backbone models.

\subsection{Experimental Setup}
\label{sec:experiment-setup}


\subsubsection{Datasets}
We use datasets that provide positive and hard-negative passage pairs. The original work of \citeauthor{reichman2024dense} used the NQ dataset, so we also experiment on this dataset. In Section~\ref{sec_repro_challenges} we have mentioned the issues related to identify which version of the NQ dataset \citeauthor{reichman2024dense} used. For our experiments, we construct a subset for experimentation as follows. We use the negative labels for the NQ dataset that are distributed in the DPR official repository\footnote{\url{https://github.com/facebookresearch/DPR}.}; specifically the files \textit{data.retriever.nq-adv-hn-train} for training and \textit{data.retriever.nq-dev} for testing. Each query--positive passage pair forms one training sample, and we randomly sample four hard negatives from the hard negative available. If fewer than four negatives exist, we oversample; if no negatives exist, we remove the query. This preprocessing results in fewer than ten removals out of ~69,000 queries. The dataset is available at \url{https://github.com/ielab/DenseRetriever-Knowledge-Acquisition}.

In addition to NQ, we also experiment with MS MARCO to study whether findings generalize. For this we acquire the dataset from \url{https://github.com/microsoft/MSMARCO-Passage-Ranking}. Our preprocessed version follows the same construction procedure we used for NQ. We make available this version of the MS MARCO dataset in the GitHub ielab repository provided above.

\subsubsection{Models}
We compare three dense retrievers with their untrained counterparts:
\begin{enumerate}[leftmargin=14pt]
    \item \textbf{DPR vs. BERT-CLS}: We examine three configurations: (a)) using DPR-query\footnote{\href{facebook/dpr-question_encoder-single-nq-base}{facebook/dpr-question-encoder-single-nq-base}} to encode both queries and passages, (b) using DPR-passage\footnote{\href{facebook/dpr-ctx_encoder-single-nq-base}{facebook/dpr-ctx-encoder-single-nq-base}} for both queries and passages, and (c)) using DPR-query for queries and DPR-passage for passages (denoted as DPR-Paired). The BERT-CLS baseline uses BERT-BASED\footnote{\href{bert-base-uncased}{bert-base-uncased}} with CLS token representation.

    \item \textbf{Contriever vs. BERT-Mean}: We compare Contriever\footnote{\href{facebook/contriever}{facebook/contriever}} with BERT using mean pooling.

    \item \textbf{ReplLlama vs. Llama}: We compare RepLlama\footnote{\href{castorini/repllama-v1-7b-lora-passage}{castorini/repllama-v1-7b-lora-passage}} with its backbone model Llama\footnote{\href{meta-llama/Llama-2-7b-hf}{meta-llama/Llama-2-7b-hf}}, both using EOS token representation.
\end{enumerate}

\subsubsection{Experimental Procedure}
We evaluate four variations of linear probing, each with a different number of passages ($N=[2,3,4,5]$ -- one of them is always positive, the remaining are hard negatives). Training hyperparameters are detailed in Table~\ref{table:training_parameters}. To fairly compare the effectiveness of linear probing layers, we use the validation set to compute accuracy after each epoch, for each setting within a epoch of 50, we select the linear-probing layer that shows highest validation accuracy and apply it on test set.

\subsection{RQ1: Reproduction with Same Settings}

\label{sec:results}

Figure~\ref{fig:cls_image} shows the accuracy trends for four different CLS-based embeddings:the BERT-BASED model, the DPR-passage, the DPR-query encoder, and DPR-paired. Despite following the setup of the original paper, our overall accuracy remains lower, indicating that reproducibility may be sensitive to experimental details that are not fully specified, e.g. the dataset construction methodology. Nevertheless, we were able to obtain similar trends reported by \cite{reichman2024dense}, that is: DPR fine-tuning does not add discriminative power to the knowledge encoded in the model. 
As the number of passages increases from two to five, the overall accuracy decreases for all encoders. Notably, BERT-CLS generally outperforms both DPR-passage and DPR-paired, suggesting that the base, not fine-tuned model captures more discriminative information.

Contrary to \cite{reichman2024dense}’s original claim that question and context encoders perform similarly, our results show that DPR-passage often underperforms not only to the BERT-CLS model but also the DPR-query model. One likely explanation for this is that the context encoder must handle significantly longer inputs and therefore compress or discard certain details that are critical for accurate passage discrimination. 

\subsection{RQ2: Reproduction on MS MARCO}
\label{sec:msmarco}
Our results from the linear probing experiments on the MS MARCO dataset generally mirror those observed on NQ, see Figure~\ref{fig:cls_msmarco}. The overall performance trends remain similar: the BERT-CLS model and the DPR-query outperform the DPR-passage and DPR-paired. However, we observe larger fluctuations across layers in MS MARCO than in NQ, which may be attributed to differences in the nature of the queries and passages across the two datasets.

\subsection{RQ3: Generalization to Different Recipes}

\begin{figure*}[ht!]
  \centering
  \includegraphics[width=\textwidth]{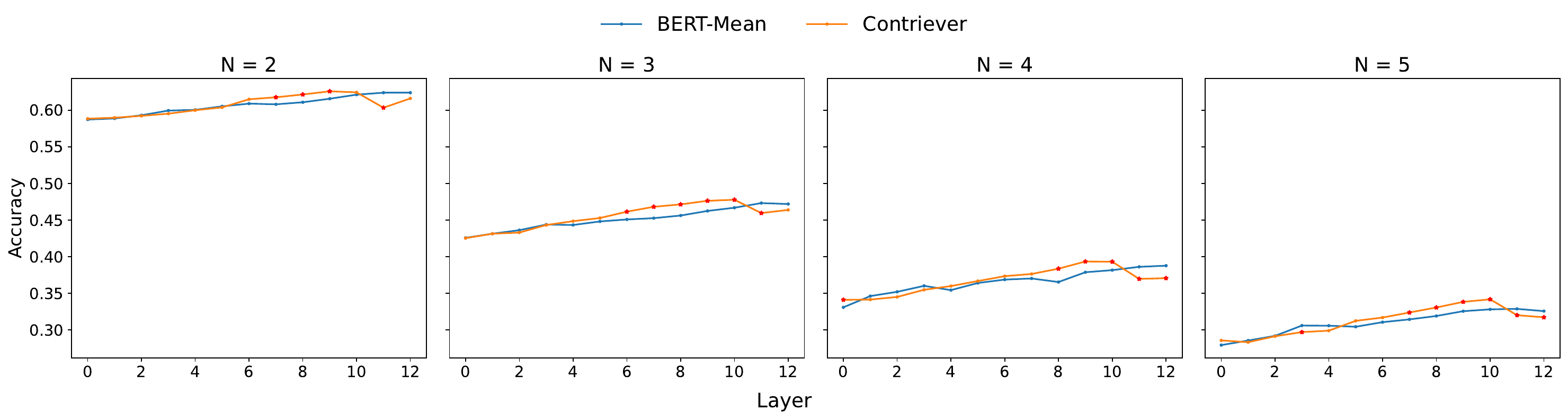}
  \caption{Generalization of Linear Probing experiment when different mean polling strategy is used (Contriever). The results are based on four different configurations for the number of passages in the probe ($N=2,3,4,5$), each represented by a sub-plot. The x-axis indicates the number of layers used for embedding learning. A two-tailed t-test ($p<0.05$) with Bonferroni correction between each DPR model with bert-based-unacsed backbone model is marked with *.}
  \label{fig:mean_image}
\end{figure*}

\begin{figure*}[ht!]
  \centering
  \includegraphics[width=\textwidth]{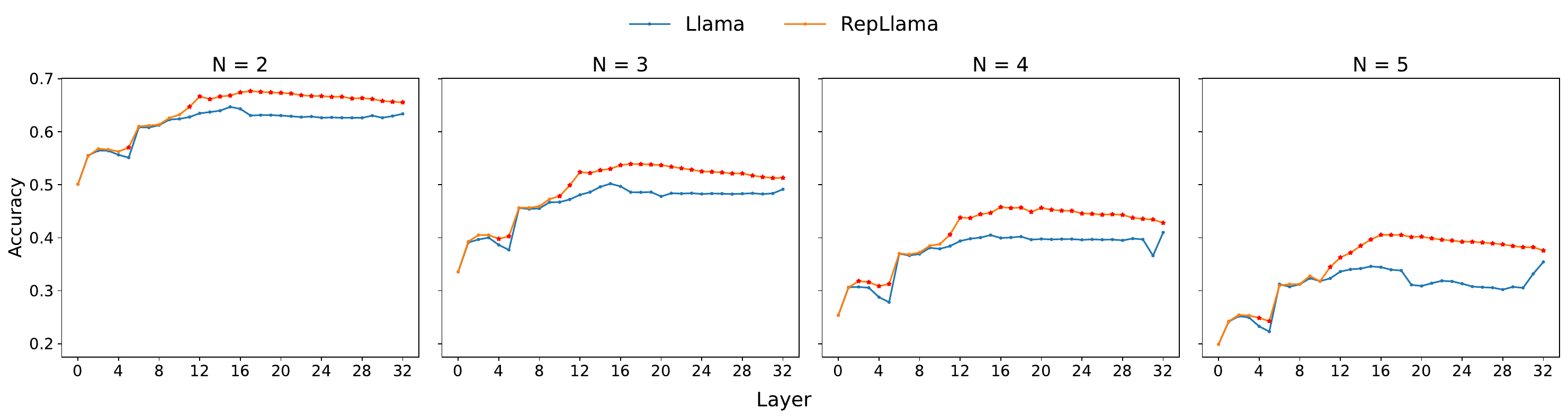}
  \caption{Generalization of Linear Probing Accuracy when decoder model backbone is used (Llama), EOS tokens are used for embedding representation. The results are based on four different configurations for the number of passages in the probe ($N=2,3,4,5$), each represented by a sub-plot. The x-axis indicates the number of layers used for embedding learning. A two-tailed t-test ($p<0.05$) with Bonferroni correction between each DPR model with bert-based-unacsed backbone model is marked with *.}
  \label{fig:eos_image}
\end{figure*}


Figure~\ref{fig:mean_image} presents the results obtained when employing mean pooling of the token embeddings for the base BERT-Mean and Contriever, a comparison not considered in the original study. In contrast to the CLS embedding experiments, the gap between the two models is notably smaller. Both achieve closer accuracy scores across layers, indicating that mean pooling may capture certain semantic cues in a more uniform way.  These findings suggest that the choice of pooling strategy can substantially influence the observed performance, sometimes overshadowing the differences among underlying model architectures or training regimes.

\subsection{RQ4: Generalization to Other Backbone}

Finally, Figure~\ref{fig:eos_image} reports the results for the EOS embedding approach applied to Llama and RepLlama. We find these to be the most striking results: Here, the performance of the RepLlama variant dramatically increases beyond the 12th layer, especially in scenarios with larger number of passages. This jump implies that deeper transformer layers may encode richer contextual signals relevant to passage discrimination, and that RepLlama in particular is able to leverage these deeper representations more effectively than Llama. The marked improvement in later layers indicates that the EOS token might encapsulate the most salient information for determining passage relevance in these models. 

\subsection{Summary}

In general, our results across these three embedding strategies highlight the following.
\begin{enumerate}[leftmargin=10pt]
    \item For the CLS embedding, due to the removal of some information, the \emph{DPR-ctx} generally has worse performance. 
    \item Different embedding pooling methods (CLS, mean, EOS) can produce divergent results, occasionally overshadowing differences between base and specialized retriever models.
    \item Model-specific variations (e.g., question vs.\ context encoders in DPR, or \emph{llama} vs.\ \emph{repllama}) can become more pronounced at deeper layers, where representational capacity is typically richer.
\end{enumerate}

Despite attempting to follow the setup described by \citeauthor{reichman2024dense}, our overall accuracy remains lower, indicating that reproducibility may be sensitive to experimental details that are not fully specified or are difficult to replicate exactly. However, these findings offer additional evidence that BERT’s pre-trained features already capture substantial passage discriminative capacity, while specialized retriever training may further enhance performance in certain configurations.

	\section{Knowledge Decentralization in Fine-tuned Dense Retrievers}

\begin{figure}[t!]
  \centering
  \includegraphics[width=\columnwidth]{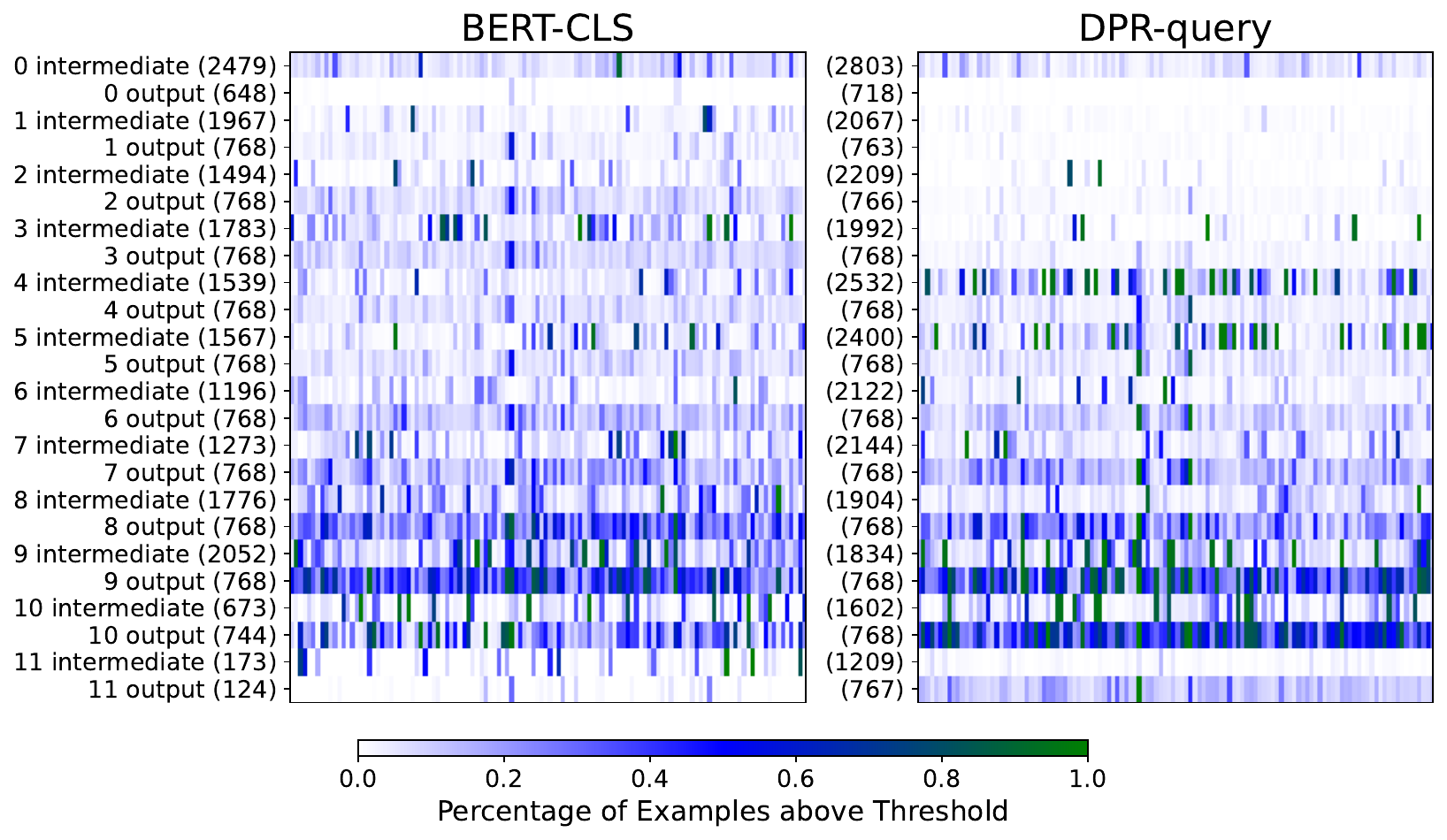}
  \includegraphics[width=\columnwidth]{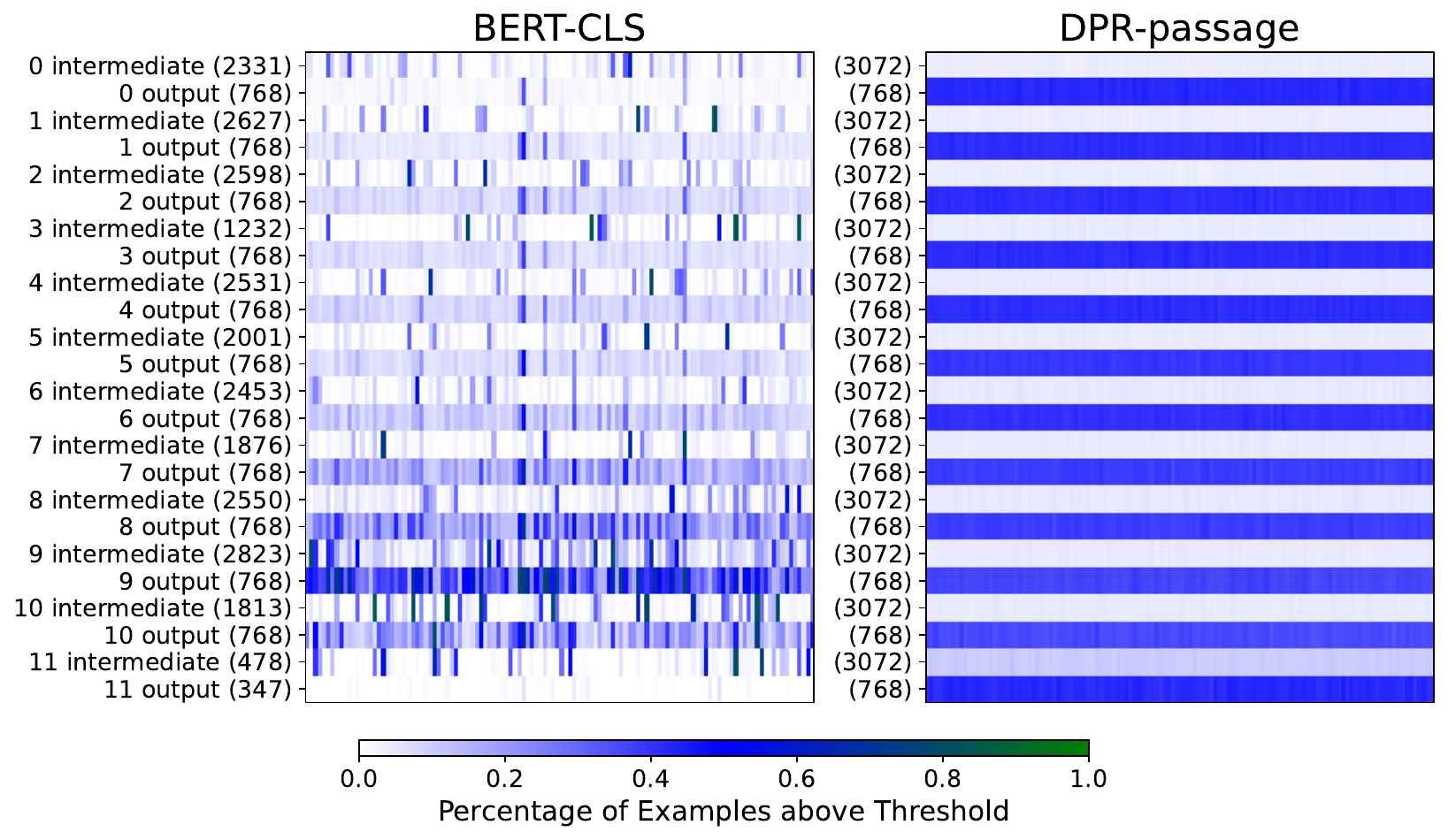}
  \caption{Results for knowledge neuron activation comparisons obtained on the NQ dataset. The plots compare DPR-query encoder vs. BERT-CLS using questions; and DPR-passage vs. BERT-CLS backbone model using positive passages.}
  \label{fig:cls-question-task2}
  
\end{figure}

\begin{figure}[t!]
  \centering
  \includegraphics[width=\columnwidth]{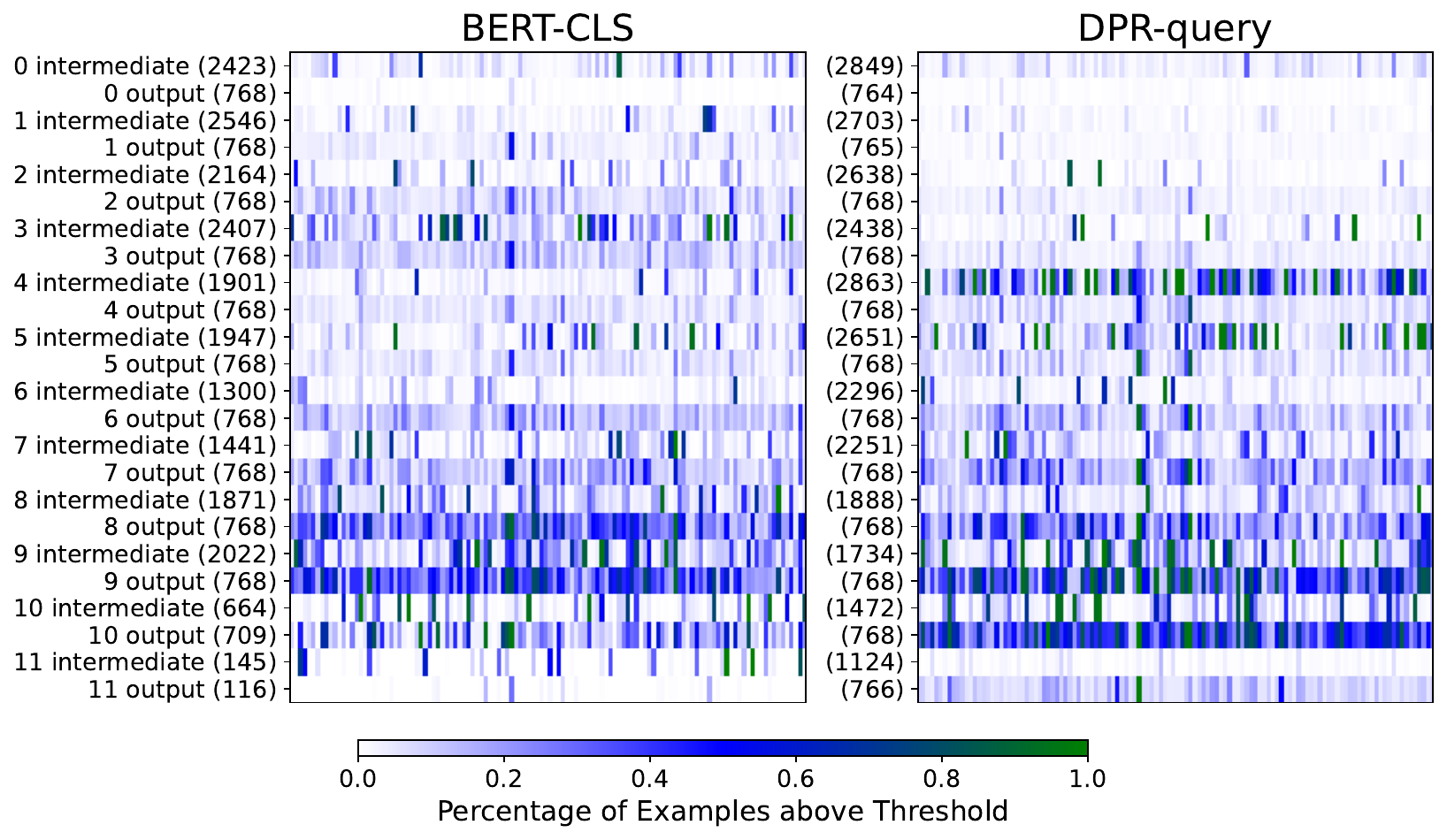}
  \includegraphics[width=\columnwidth]{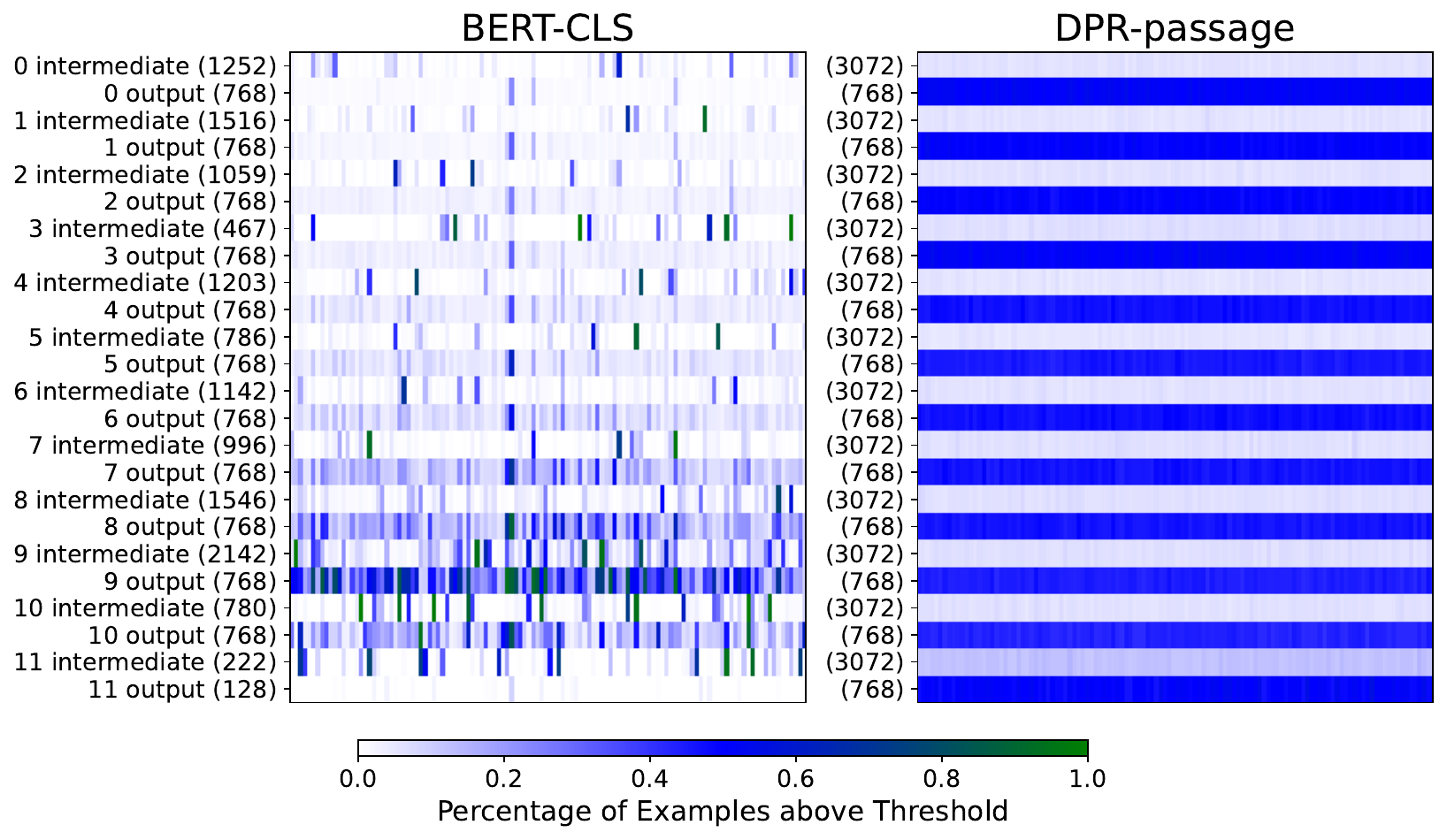}
  \caption{Results for knowledge neuron activation comparisons obtained on the MS MARCO dataset. The plots compare DPR-query encoder vs. BERT-CLS using questions; and DPR-passage vs. BERT-CLS using positive passages.}
  \label{fig:cls-msmarco-task2}
\end{figure}

\begin{figure}[t!]
  \centering
  \includegraphics[width=\columnwidth]{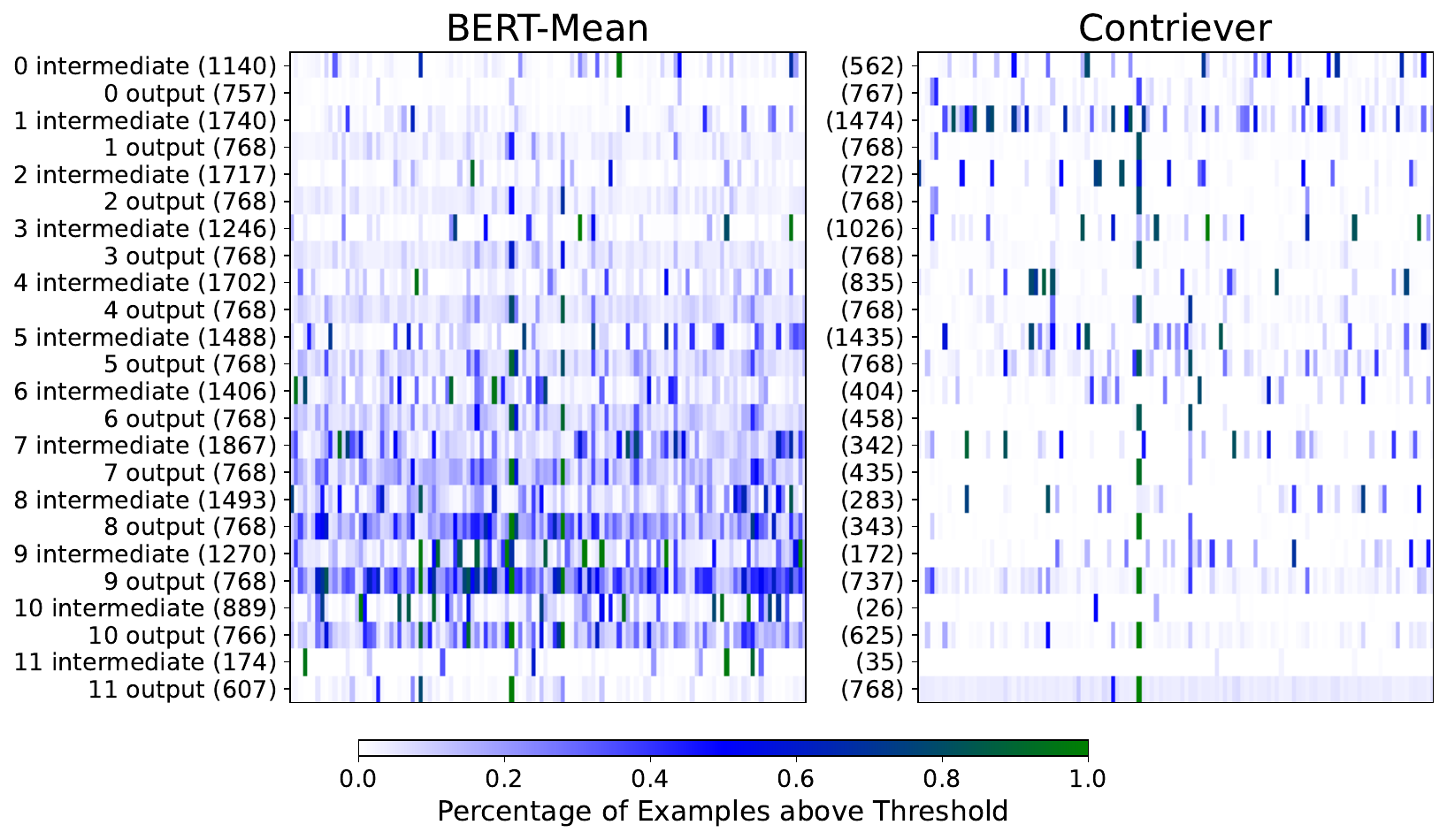}
  \caption{Results for knowledge neuron activation comparisons obtained on the NQ dataset. The plots compare Contriever vs. BERT backbone model using Mean-embedding representation.}
  \label{fig:mean-task2}
\end{figure}

\begin{figure}[ht!]
  \centering
  \includegraphics[width=\columnwidth]{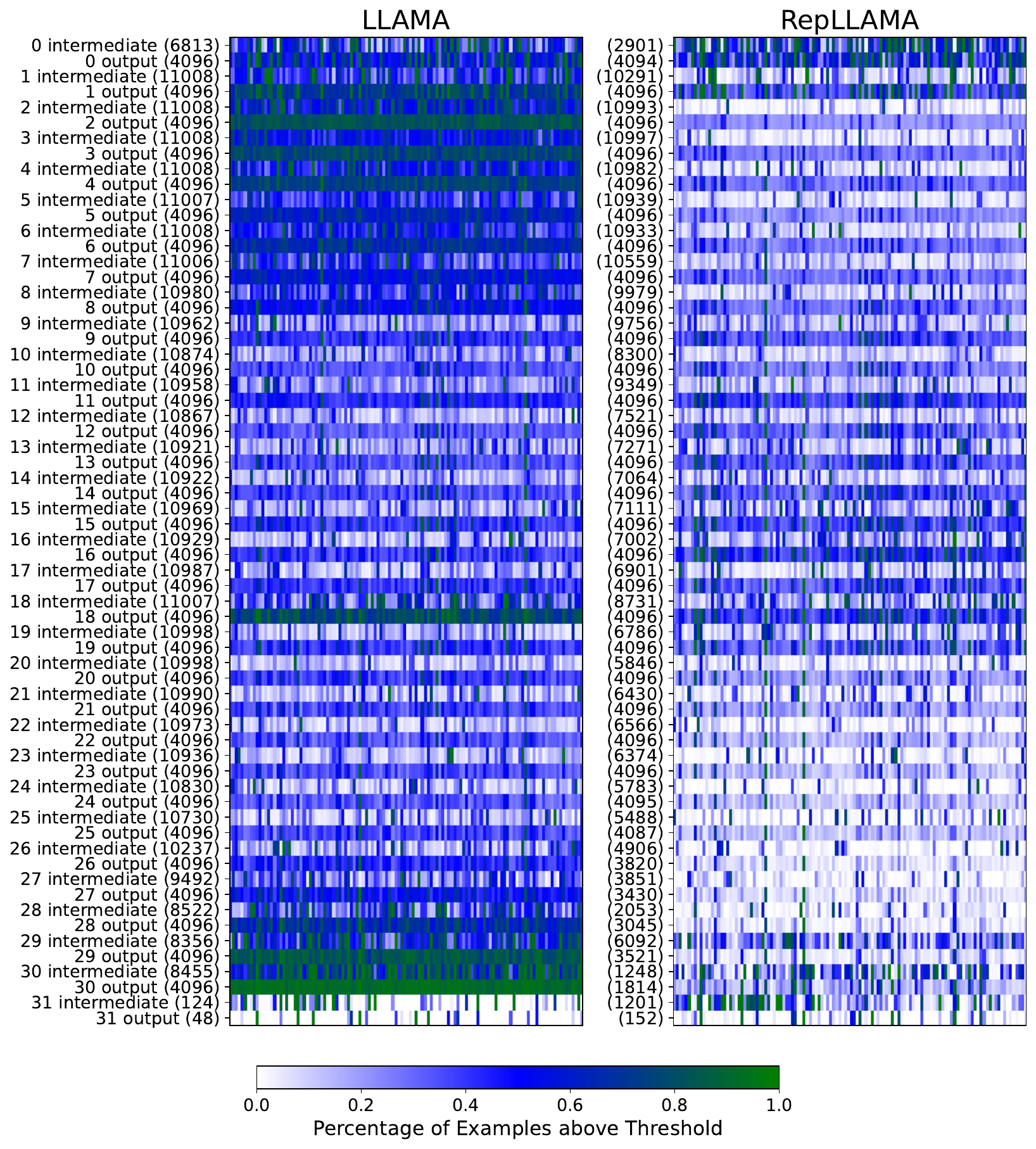}
  \caption{Results for knowledge neuron activation comparisons obtained on the NQ dataset. The plots compare Repllama vs. Llama backbone model using EOS token representation. }
  \label{fig:eos-question-task2}
\end{figure}

Next, we investigate the phenomenon of knowledge decentralization in fine-tuned dense retrievers by synthesizing insights from a neuron attribution analysis, following the original methodology of~\citet{reichman2024dense}. This analysis aims to measure the contribution of individual neuron activations to the final embedding of the model, providing a structured way to assess how knowledge is distributed across the network.

To achieve this, we employ an integrated gradient-based (IG) attribution technique~\citep{Dai2022}, which quantifies the impact of each neuron's activation on the final representation. By applying this method, we examine whether fine-tuning redistributes knowledge within the model, shifting from a more centralized to a decentralized retrieval strategy. Specifically, we explore whether fine-tuning leads to increased activation in intermediate layers, allowing multiple neurons to contribute to knowledge retrieval rather than relying on a few highly activated neurons. The following subsections present our experimental results, analyzing knowledge decentralization across different architectures.

\subsection{Methodology: Neuron Activation Analysis}
\label{sec:methodology-task2}

Neuron activation analysis consists of two stages: (1) \emph{computing neuron attributions via integrated gradients} and (2) \emph{applying a threshold to identify active neurons and observe global patterns}.

\paragraph{Stage 1: Integrated Gradients Computation.}
We focus on the linear sub-layers (\emph{intermediate} and \emph{output} dense layers) within each Transformer block, as these layers have been argued to store much of the model’s factual knowledge~\citep{Geva2021}. Let $w_i^{(l)} \in \mathbb{R}^{d}$ be the parameter vector for the $i$-th neuron in layer $l$. To compute the neuron attribution, we follow the standard integrated gradients approach~\citep{Sundararajan2017,Hao2021}, varying $w_i^{(l)}$ from $\mathbf{0}$ to its learned value:

\begin{equation}
    \mathrm{Attr}\bigl(w_i^{(l)}\bigr) = \int_{0}^{1} 
    \frac{\partial\, \mathbf{P}_x\!\bigl(\alpha\, w_i^{(l)}\bigr)}{\partial\, w_i^{(l)}} 
    \,\mathrm{d}\alpha,
    \label{eq:ig-attr}
\end{equation}

where $\mathbf{P}_x(\cdot)$ denotes the model’s scalar output for input $x$, \emph{holding all other parameters fixed} while varying only the neuron’s weight vector. Since the continuous integral is intractable, we approximate it using a Riemann sum over discrete integration steps.

\paragraph{Stage 2: Thresholding and Aggregation.}
After computing neuron attributions across all examples, we \emph{normalize} each attribution by the maximum observed value within the same example across the entire network. Following~\citet{reichman2024dense}, we apply a default threshold of $0.1 \times \max(\mathrm{Attr})$ within each example, yielding a coarse selection of ``active'' neurons. We then aggregate these counts across the dataset to observe the frequency with which each neuron exceeds this threshold.
This methodology enables a direct comparison of how neuron activation patterns shift before and after fine-tuning, revealing which linear sub-layer neurons play a dominant role in producing the final representations.

\subsection{Experimental Setup}
\label{sec:experiment-setup-task2}

In this section, we describe how we apply the neuron activation analysis to the same model architecture comparisons and datasets introduced in Section~\ref{sec:experiment-setup}.

\subsubsection{Datasets}
We use the same NQ and MS MARCO test subsets described in Section~\ref{sec:experiment-setup}, focusing only on the questions (for the query encoder) and positive passages (for the passage encoder). Since our objective is to analyze neuron attribution activation patterns rather than evaluate retrieval effectiveness, each question or passage is processed independently.

\subsubsection{Models}
We conduct neuron activation analysis on the same set of fine-tuned dense retriever and their corresponding untrained backbone model pairs as in the previous experiments:
\begin{enumerate}[leftmargin=10pt]
    \item \textbf{DPR vs.\ BERT-CLS}: We analyze neuron activations in both the DPR models (\textit{query} and \textit{passage}) and the BERT-CLS baseline.
    \item \textbf{Contriever vs.\ BERT-Mean}: We compare neuron activations in Contriever and the mean-pooled BERT baseline.
    \item \textbf{ReplLlama vs.\ Llama}: We study neuron activations in ReplLlama and its corresponding Llama backbone, using the EOS token representation.
\end{enumerate}

\subsubsection{Experimental Procedure}
We apply the IG computation method described in Section~\ref{sec:methodology-task2} to each model’s linear sub-layers (i.e., the \emph{intermediate} and \emph{output} dense layers in each Transformer block). For each input text (either a question or passage), we pass the tokenized input through the model and compute the IG attributions for all neurons by holding the rest of the parameters fixed and integrating each neuron’s weights from \(\mathbf{0}\) to their learned values. We set $n_{\text{steps}}=20$, consistent with the original IG computation methodology~\cite{Dai2022}. To quantify neuron activation, we normalize attributions by the maximum value within each example and count the number of neurons that exceed a threshold of \(0.1 \times \max(\mathrm{Attr})\). We then aggregate these counts across the dataset to assess how frequently each neuron is \emph{active} for different inputs. 

Unlike the original work of \citet{reichman2024dense}, which reports the absolute number of examples activating each neuron -- potentially leading to inconsistencies across datasets -- we instead compute the \emph{percentage} of examples exceeding the threshold. This normalization ensures comparability across datasets and model variants, providing a more robust analysis of neuron activation patterns.

\subsection{RQ1: Reproduction with Same Settings}

Figure~\ref{fig:cls-question-task2} compares the neuron activation patterns in the intermediate and output layers of \emph{DPR-query} vs. \emph{BERT-CLS}, and \emph{DPR-passage} vs. \emph{BERT-CLS} when tested on the NQ dataset, following the original investigation by~\citet{reichman2024dense}.

The original study conceptualized the intermediate layers as ``keys'' for accessing internally stored semantic knowledge, while the output layer serves as a ``value'' layer, representing the encoded text knowledge. It was observed that DPR fine-tuning leads to increased neuron activations in the intermediate layers and fewer activations in the early output layers. This suggests that instead of relying on the text explicitly encoded during training, the DPR-fine-tuned model primarily accesses its internal knowledge. Consequently, DPR fine-tuning does not modify the internally stored knowledge of the pre-trained model but rather alters how it is accessed.

Our reproduction under the same experimental setting confirmed the findings of \citeauthor{reichman2024dense} when comparing \emph{DPR-query} with \emph{BERT-CLS}. Specifically, we observed increased neuron activations in the intermediate layers, while the early output layers exhibited fewer activated neurons.

However, when we compare \emph{DPR-passage} with \emph{BERT-CLS}, we find a completely different trend. While DPR-tuning does increase neuron activations in the intermediate layers, the output layer maintains consistent importance across all layers, with a higher attribution in the later layers.


\subsection{RQ2: Reproducing on MS MARCO}

Next, we examine whether the decentralization of knowledge observed in DPR training is specific to the NQ dataset. To do this, we conduct the same analysis using the MS MARCO dataset. Figure~\ref{fig:cls-msmarco-task2} presents the neuron activation patterns for \emph{DPR-query} vs. \emph{BERT-CLS} and \emph{DPR-passage} vs. \emph{BERT-CLS} when tested on MS MARCO.

Our results reveal the same trends observed in the NQ dataset: \emph{DPR-query} exhibits increased neuron activations in the intermediate layers while showing fewer activated neurons in the early output layers, whereas \emph{DPR-passage} maintains consistent activation across all layers, with stronger attribution in the later output layers. This pattern mirrors our findings from the NQ dataset, reinforcing that DPR fine-tuning affects queries and passages differently.


\subsection{RQ3: Generalization to Different Recipe}

We then investigate whether the findings from DPR models transfer to Contriever tuning and mean pooling. We present the neuron activation pattern results for this comparison in Figure~\ref{fig:mean-task2}. 

Contrary to our observations made for DPR models, we find that \emph{Contriever} consistently exhibits fewer activated neurons across all intermediate layers compared to its backbone model. One possible explanation for this is that mean pooling inherently distributes the representation across all tokens in the sequence, reducing the need for highly specialized activations in intermediate layers. As a result, the model may rely more on distributed encoding rather than activating specific neurons to capture semantic knowledge.

Furthermore, we observe that the output layer of \emph{Contriever} does not show the same early-layer suppression seen in DPR models. Instead, it maintains a more uniform activation pattern, suggesting that the retrieval mechanism in \emph{Contriever} differs fundamentally from DPR in how it accesses and structures knowledge.

These findings indicate that while DPR fine-tuning decentralizes knowledge retrieval through increased intermediate-layer activations, \emph{Contriever} relies on a more uniform distribution of activation across layers. This distinction suggests that different retrieval architectures employ varied mechanisms for encoding and retrieving knowledge. Moreover, using tuning architectures such as mean pooling appears to influence how knowledge is structured and accessed within the model. Specifically, mean pooling may encourage a more evenly spread representation across tokens, reducing reliance on localized neuron activations. This observation raises further questions about how different pooling strategies impact retrieval effectiveness and whether certain architectures inherently prioritize different aspects of semantic encoding.

\subsection{RQ4: Generalization to Other Backbone}

Finally, we investigate whether the findings hold when using a decoder-based dense retriever backbone, specifically \emph{Llama}. Figure~\ref{fig:eos-question-task2} displays the neuron activation patterns for \emph{Llama} and \emph{RepLlama}. 

From our results, \emph{RepLlama} exhibits fewer active neurons in the intermediate layers compared to its backbone counterpart\footnote{Note we set the activation threshold to 0.01 for Llama-based models, as this threshold provided the clearest trend, similar results were observed across other threshold settings; results for these settings are included in our Github repository.}.

This finding suggests that decoder-based architectures may follow a different retrieval mechanism compared to DPR-based models. The reduction in intermediate-layer activations in \emph{RepLlama} could indicate a more compressed or distributed knowledge representation, where the model relies on fewer but more specialized neurons for retrieval. This contrasts with the decentralization trend observed in DPR-query models and highlights the need for further exploration into how different architectural choices impact knowledge access and retrieval efficiency.

\subsection{Summary}

Overall, our findings highlight the diverse ways in which different retrieval architectures structure and access knowledge. While DPR fine-tuning decentralizes knowledge retrieval by increasing intermediate-layer activations in query models, passage models exhibit a different trend, maintaining consistent activation across layers. This pattern remains consistent across datasets, indicating that DPR's impact is not data-dependent. 

In contrast, however, Contriever, which employs mean pooling, demonstrates a more uniform activation distribution, suggesting an alternative mechanism for knowledge retrieval. Also decoder-based retrievers, such as \emph{RepLlama}, show reduced intermediate-layer activations, indicating a potentially more compressed knowledge representation. These results emphasize the need for further research to understand how architectural choices influence retrieval behavior and efficiency. We also acknowledge that the method the original authors used for investigating knowledge attribution, and that we also relied upon in our reproduction (i.e., neuron attribution), though stemming from previous literature it may actually be not reliable for investigating dense retrieval. 

	\section{Related Works}

Dense retrieval relies on deep neural networks to encode textual data into dense vector representations, enabling efficient approximate nearest neighbor search. These models are broadly categorized into encoder-based and decoder-based retrievers, each employing different representation and retrieval strategies.

\textbf{Encoder-based dense retrievers}, such as DPR~\cite{karpukhin2020dense}, utilize transformer encoders to map queries and documents into fixed-size vectors. The retrieval process is then performed by computing the dot product between query and document representations. One common approach is the use of the \textbf{[CLS] token representation}, where the final hidden state of the special [CLS] token is extracted as a compact representation of the entire input sequence. While effective, this method has been observed to focus more on the beginning of the input, potentially missing finer-grained information distributed throughout the text.

An alternative strategy is \textbf{mean pooling}, as adopted by Contriever~\cite{izacard2021contriever}, where the final embeddings of all tokens are averaged to form a unified document representation. This pooling mechanism captures a more distributed representation of the input and is often preferred in cases where information is spread across longer passages. Furthermore, some models, such as Sentence-BERT~\cite{reimers2019sentence} and SimCSE~\cite{gao2021simcse}, extend dense retrieval capabilities to zero-shot settings, leveraging contrastive learning and pre-trained embeddings to provide robust document representations without requiring task-specific fine-tuning.

\textbf{Decoder-based dense retrievers} are in contrast to encoder-based models, incorporating autoregressive decoding mechanisms for retrieval. RePLAMA~\cite{replama2021} exemplifies this approach by reframing the retrieval task as a sequence generation problem, where the model generates candidate passages based on query context rather than performing direct similarity matching. This method enhances retrieval flexibility by capturing long-range dependencies and richer query-document relationships. Similarly, PromptReps~\cite{promptreps2021} integrates prompt-based learning with dense retrieval, employing carefully designed prompts to guide the model in generating more discriminative representations. These approaches illustrate the growing shift toward generative retrieval frameworks that combine aspects of traditional retrieval with neural sequence modeling.

\textbf{Benchmarks} such as \textbf{MS MARCO Passage Ranking}~\cite{msmarco} and \textbf{Natural Questions (NQ)}~\cite{naturalquestions} are commonly used to evaluate Dense retrievers.
 MS MARCO consists of web search queries with passage relevance annotations, while NQ features real-world questions paired with relevant Wikipedia passages. These datasets provide diverse and realistic challenges, making them essential for assessing retrieval effectiveness and advancing research in dense retrieval.

\section{Discussion and Conclusion}
\label{sec:conclusion}

Understanding knowledge processing and knowledge flows in dense retrieval models is essential for identifying potential methodological gaps and new directions, e.g., establishing new masked language model training stages after fine-tuning for more effective dense retrieval. Yet, these aspects are rarely investigated. 

In this paper we considered previous work by \citeauthor{reichman2024dense} who questioned the role of fine-tuning vs. that of pre-training within dense retrievers~\cite{reichman2024dense}. In particular, we aimed to reproduce and validate two core claims emerging from their work: (1) that pre-trained language models already possess strong discriminative capacity for determining passage relevance, and (2) that dense retrieval training decentralizes knowledge by activating broader neuron pathways. To do so, we carried out extensive experimentation considering additional training recipes, representation methods, backbones and datasets. 


\textbf{Key Findings.} Our results lead to the following observations:
\begin{itemize}[leftmargin=14pt]
	\setlength\itemsep{3pt}
    \item \textbf{Discriminative Capacity:} Linear probing experiments confirmed that pre-trained BERT models achieve 50–60\% accuracy in distinguishing relevant passages, with DPR fine-tuning yielding comparable performance. This aligns with \citeauthor{reichman2024dense}’s argument that pre-trained knowledge, rather than fine-tuning, primarily governs retrieval effectiveness. Similar trends were observed in Contriever fine-tuning and mean-pooling. However, decoder-based dense retrievers (e.g., ReplLlama) using EOS token pooling exhibited an 18–22\% improvement in deeper-layer accuracy, suggesting that architecture and embedding strategies play a significant role in how knowledge is altered during dense retrieval training.
    \item \textbf{Knowledge Decentralization:} Integrated gradient analysis showed that DPR fine-tuning increases intermediate-layer activations by 32–41\%, supporting the decentralization hypothesis proposed by \citeauthor{reichman2024dense}. However, this effect reversed when using mean-pooling (Contriever) or an EOS-token-based decoder (Llama), where fine-tuning instead \emph{reduced} neuron activation breadth. These results suggest that knowledge decentralization is dependent on the model’s architecture and pooling strategy.
    \item \textbf{Reversed Trends in DPR-Passage Encoder:} Our experiments revealed differing trends between question and passage encoders: (1) both the DPR-question encoder and the BERT backbone exhibited significantly higher discriminative power than the passage encoder, and (2) knowledge neurons were more activated in the output layer of the passage encoder. These findings contradict the original claim by \citeauthor{reichman2024dense}, suggesting that passage encoders may rely on different retrieval dynamics than query encoders.
\end{itemize}

\textbf{Implications.}
Our findings suggest that the conclusions by \citeauthor{reichman2024dense}~\cite{reichman2024dense} are architecture- and task-dependent. While BERT-based DPR models exhibit knowledge decentralization as originally described, this behavior does not generalize universally -- Llama-based retrievers and mean-pooled models exhibit distinct activation patterns. These results emphasize the need to evaluate retrieval mechanisms across diverse architectures rather than relying on BERT-centric assumptions.

\textbf{Limitations and Future Work.}
Our replication study encountered four primary limitations: (1) Lower accuracy compared to the original results—our values are 10–15\% lower in absolute magnitude—likely due to differences in datasets and hyperparameters. As the original data setup details were unavailable, this discrepancy is expected. (2) We did not investigate additional model architectures, such as encoder-decoder models (e.g., T5) or models with more than 7B parameters. (3) We provide only preliminary explanations rather than a comprehensive analysis of why the findings obtained by the original authors on DPR (and successfully reproduced by us) do not generalize to the other architectures considered in our paper. (4) Our analysis primarily relies on techniques used in the original paper we reproduce, which are established techniques in field—such as linear probing and neuron attribution—to assess the presence of knowledge in models. We did not critically evaluate or validate the reliability of these methods, and their suitability to dense retrieval.

 Future work should: (1) Extend analyses to other language model backbones and sizes, and (2) Investigate why certain pooling strategies (e.g., EOS) amplify fine-tuning benefits.

Ultimately, our study reinforces the notion that in DPR models, pre-trained knowledge serves as the foundation for retrieval performance, while fine-tuning plays a more nuanced role—primarily modifying neuron activation patterns rather than fundamentally reorganizing knowledge storage. However, this trend does not generalize across all architectures. In contrast, models employing mean pooling (e.g., Contriever) or decoder-based retrieval (e.g., Llama) exhibit different behavior, where fine-tuning may reduce neuron activation breadth rather than decentralizing knowledge. These insights highlight the need for deeper mechanistic analyses of retrieval models across diverse architectures to understand how different training paradigms influence knowledge access and representation. Future research could explore targeted architecture modifications to improve knowledge flow in dense retrievers. Our code and results are made available at~\url{https://github.com/ielab/DenseRetriever-Knowledge-Acquisition}.

	\bibliographystyle{plainnat}
	\balance
	\bibliography{references}
	
\end{document}